\documentclass[review,sort,compress]{elsarticle}
\usepackage{lineno,hyperref}
\modulolinenumbers[5]

\usepackage[utf8]{inputenc}
\usepackage{graphicx}
\usepackage[normalem]{ulem}
\usepackage{tabularx}
\usepackage{todonotes}
\usepackage{url}
\usepackage{color,soul}
\usepackage{caption}
\usepackage{amsmath,amssymb,mathtools}
\usepackage{color}
\usepackage{algorithm}
\usepackage{algorithmic}

\DeclareMathOperator{\E}{\mathbb{E}}

\usepackage[bottom]{footmisc}
\usepackage{url}

\author[NW_address1]{Zijiang Yang \corref{ec1}}
\ead{zijiangyang2016@u.northwestern.edu}
\cortext[ec1]{These authors contributed equally to this work.}

\author[NW_address2]{Xiaolin Li \corref{ec1}}
\ead{xiaolinli2018@u.northwestern.edu}

\author[Duke_address]{L. Catherine Brinson}
\ead{cate.brinson@duke.edu}

\author[NW_address1]{Alok N. Choudhary}
\ead{a-choudhary@northwestern.edu}

\author[NW_address3]{Wei Chen \corref{cor1}}
\ead{weichen@northwestern.edu}

\author[NW_address1]{Ankit Agrawal \corref{cor1}}
\ead{ankitag@eecs.northwestern.edu}
\cortext[cor1]{Corresponding authors, ankitag@eecs.northwestern.edu, or weichen@northwestern.edu}

\address[NW_address1]{Department of Electrical Engineering and Computer Science, Northwestern University, Evanston, IL, 60208 USA}

\address[NW_address2]{Theoretical and Applied Mechanics, Northwestern University, Evanston, IL, 60208 USA}

\address[NW_address3]{Department of Mechanical Engineering, Northwestern University, Evanston, IL, 60208 USA}

\address[Duke_address]{Department of Mechanical Engineering and Materials Science, Duke University, Durham, NC, 27708 USA}

\journal{JMD special issue on ``Design of Engineered Materials and Structures''}
\begin{document}
\begin{frontmatter}

\title{Microstructural Materials Design via Deep Adversarial Learning Methodology}

\begin{abstract}
Identifying the key microstructure representations is crucial for Computational Materials Design (CMD). However, existing microstructure characterization and reconstruction (MCR) techniques have limitations to be applied for microstructural materials design. Some MCR approaches are not applicable for microstructural materials design because no parameters are available to serve as design variables, while others introduce significant information loss in either microstructure representation and/or dimensionality reduction. In this work, we present a deep adversarial learning methodology that overcomes the limitations of existing MCR techniques. In the proposed methodology, generative adversarial networks (GAN) are trained to learn the mapping between latent variables and microstructures\footnote{The GAN model is available for download at: \url{https://github.com/zyz293/GAN_Materials_Design}}. Thereafter, the low-dimensional latent variables serve as design variables, and a Bayesian optimization framework is applied to obtain microstructures with desired material property. Due to the special design of the network architecture, the proposed methodology is able to identify the latent (design) variables with desired dimensionality, as well as capturing complex material microstructural characteristics. The validity of the proposed methodology is tested numerically on a synthetic microstructure dataset and its effectiveness for microstructural materials design is evaluated through a case study of optimizing optical performance for energy absorption. Additional features, such as scalability and transferability, are also demonstrated in this work. In essence, the proposed methodology provides an end-to-end solution for microstructural materials design, in which GAN reduces information loss and preserves more microstructural characteristics, and the GP-Hedge optimization improves the efficiency of design exploration.
\end{abstract}

\begin{keyword}
Microstructural materials design \sep Microstructural analysis \sep Deep learning \sep Generative adversarial network \sep Bayesian optimization \sep Scalability \sep Transfer learning
\end{keyword}

\end{frontmatter}

\section{INTRODUCTION}
\begin{figure}[!ht]
\centering
\includegraphics[width = 4.5in, keepaspectratio=true]{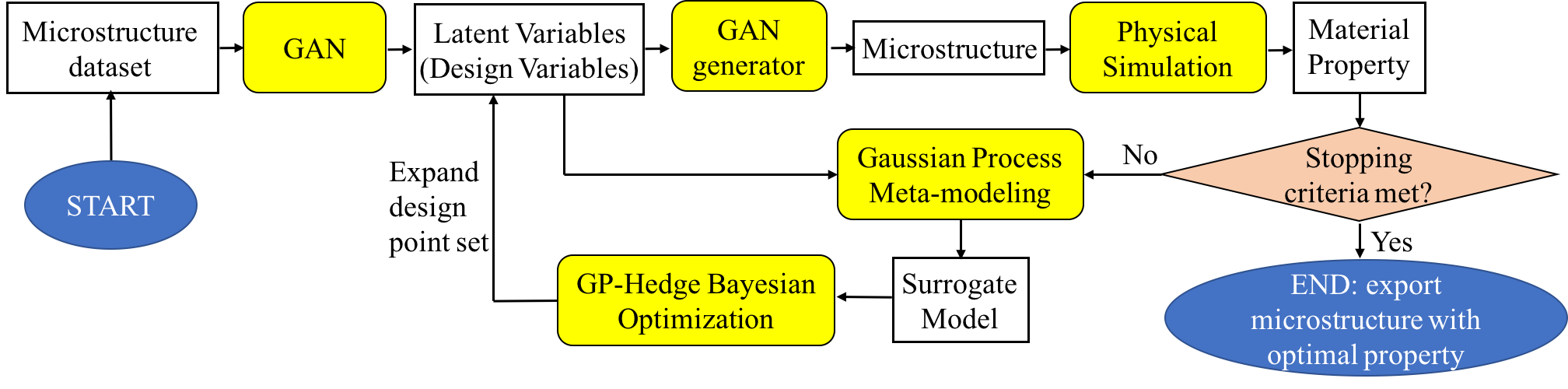}
\caption{The flowchart of the proposed design methodology.}
\label{flowchart}
\end{figure}

To date, Computational Materials Design (CMD) has revolutionarily changed the way advanced materials are developed \cite{thornton2009computational, olson1997computational, saito2013computational, kuehmann2009computational, vickers2015materials, agrawal2016perspective,zhao2016perspective,wang2018Adaptive}. In the plethora of successes in CMD \cite{paul2008polymer, natarajan2013effect, hassinger2016toward, brough2017microstructure,liu2017context,yang2017deep,yu2017characterization}, microstructure sensitive design \cite{fullwood2010microstructure} has shown its significance in driving the rapid discovery and manufacturing of new materials. In designing material microstructures, the appropriate design representation of microstructures determines its ultimate success. A common practice of selecting microstructural design variables is to choose key microstructure characteristics from existing microstructure characterization and reconstruction techniques (MCR). A comprehensive review of existing MCR techniques is provided by Bostanaband et al. \cite{bostanabad2018computational}. Together with some recent works using deep learning, the existing techniques are classified into the following categories:\\
1. Correlation function-based methods \cite{jiao2007modeling}\\
2. Physical descriptor-based methods \cite{xu2014descriptor}\\
3. Gaussian Random Field (GRF)-based methods \cite{jiang2013efficient}\\
4. Markovian Random Field (MRF)-based methods \cite{bostanabad2016stochastic}\\
5. Deep Belief Network-based methods \cite{cang2017microstructure}\\
6. Spectral Density Function (SDF)-based methods \cite{yu2017characterization}, and\\
7. Transfer Learning-based methods \cite{li2018transfer, lubbers2017inferring}\\
\indent However, not all existing MCR techniques are applicable for microstructural materials design. Two major limitations exist: 1) Some MCR methods (methods 3, 4, 5 and 7) are not applicable for microstructural materials design, because no parameters are available to serve as design variables for generating new microstructure designs. 2)  While methods 1, 2 and 6 are applicable for microstructural materials design, their efficacy is limited by the potential information loss (i.e. loss of either dispersive or geometrical characteristics) in microstructure representation and/or dimensionality reduction. In microstructure representations, some approximations such as taking radial averages in method 1 \& 6 or approximating cluster shapes with ellipses in method 2 could result in the loss of microstructural characteristics. Dimension reduction is often needed in microstructure optimization due to the high-dimensional representation of microstructures. A common practice is to conduct a transformation of microstructure representations (e.g. using Principal Component Analysis (PCA)) and remove some insignificant dimensions. Information loss would also occur in the removal process. For instance, Paulson et al.~\cite{paulson2017reduced} use spatial correlation function as the microstructure representation, and conduct a PCA transformation. It is shown in their work that removing some principal components could lead to a significant reduction in explained structural variance. Another example is the use of descriptor-based approach. After obtaining the full list of descriptors, a supervised learning-based feature selection step is often used to remove the lower-ranked descriptors~\cite{xu2015machine}, wherein some geometric or higher-order dispersive information is lost. It should be noted that the aforementioned dimensionality reduction techniques do not guarantee the capability of generating new microstructural designs using the reduced dimension. For example, while the principal components learned by PCA are capable of identifying new dimensions that are not linearly correlated, it is not clear how to generate a new microstructure by sampling in the learned principal dimensions.\\
\indent Compared to the existing MCR techniques, generative models are promising alternatives to address the problems in microstructural materials design. Instead of identifying characteristics from microstructures, generative models emphasize the ability of using a low-dimensional latent variables $Z$ to generate high-dimensional data $X$ through a generative mapping $G : Z \rightarrow X$ to approximate the real data probability density $P_{data}(x)$. In other words, the evaluation criteria for generative models is whether it is capable of producing very realistic samples, which are indistinguishable from real samples. The latent variables learned in the generative model can therefore serve as design variables for microstructural materials design. In addition, generative models are especially powerful for microstructural materials design because the approach is model-based and it can rapidly generate new microstructures by changing the values of latent variables, while existing MCR approaches often need tedious optimization for microstructure reconstructions (e.g. Simulated Annealing is used in correlation function-based reconstruction).\\
\indent In the realm of deep learning, Variational Auto-Encoder (VAE) \cite{kingma2013auto} and Generative Adversarial Networks (GAN) \cite{goodfellow2014generative, makhzani2015adversarial} are two major categories of generative models. It is well recognized that VAE suffers from the issue of ``maximum likelihood training paradigm'' when combined with a conditional independence assumption on the output given the latent variables, and they tend to distribute probability mass diffusely over the data space and generate blurry samples \cite{dumoulin2016adversarially}. Despite these theoretical disadvantages, both Cang et al.\cite{cang2018arxiv} and Guo et al.~\cite{guo2018indirect} developed VAE-based models for representing sandstone material microstructures and topology optimization respectively. However, their generative capability is bottlenecked at images of size $40\times40$, and it is impossible to scale up because fully-connected layers are involved in their network architecture.\\
\indent In contrast to VAE, GAN is a better choice to bypass these problems. Different from VAE, GAN identifies the latent variables of data by training a generator-discriminator model pair in adversarial manner. In \cite{mosser2017reconstruction, mosser2017stochastic}, GAN is used for reconstructing different types of microstructures, but their applications in computational materials design are unexplored. In this work, as illustrated in Figure~\ref{flowchart}, we apply a fully scalable GAN-based approach to determine the latent variables of a set of microstructures once its dimensionality is pre-specified. The latent variables are then treated as design variables in microstructure optimization. Thereafter, the material property for the latent variables is obtained by propagating the latent variables through the generator in GAN, followed by physical simulations of structure-property or structure-performance relations. Considering that physical simulations are usually computationally costly, we also want to minimize the number of property evaluations. Therefore, we pursue a response surface model-based GP-Hedge Bayesian optimization framework to optimize microstructure with desired material property/performance.\\

\indent The proposed deep adversarial learning methodology provides an end-to-end solution that offers a low-dimensional and non-linear embedding of microstructures for microstructural materials design. Compared with the existing methods which cannot fully capture microstructural characteristics (e.g. two-point correlation function in method 1 and physical descriptors in method 2), the proposed method does not make any geometrical or dispersive approximations and thus there is no information loss. In addition, the non-linear embedding of microstructures in the proposed method avoids the removal of insignificant dimensions of microstructure representations (e.g. physical descriptors in method 2 and principal components in method 1) so that more microstructural information is preserved. Moreover, the proposed method is also beneficial for microstructural materials design because the dimensionality of latent variables can be pre-specified as needed.  Meanwhile, since the GAN is implemented by deep neural networks with large model capacity, it is able to capture very complex microstructural characteristics. In addition to the contribution of the proposed approach to microstructural materials design, we also demonstrate that the proposed approach is advantageous in: 1) \textit{scalability}: the proposed approach is capable of converting microstructures into reasonable and computationally affordable low-dimensional representations as needed, and the generator in proposed model is scalable to produce arbitrary sized microstructures; 2) \textit{transferability}: the discriminator in the proposed approach could be reused to serve as a pre-trained model to facilitate the development of structure-property predictive models. To the best of the authors’ knowledge, this work is the first that applies adversarial learning in computational design of materials microstructure.\\
\indent In the remainder of this paper, we break our presentation of the deep adversarial learning design methodology into five sections. In the first part (Section 2 – Design Representation), we present the technical fundamentals of the deep adversarial learning approach, and show how the latent variables of microstructures are learned using the proposed approach. The latent variables are then treated as design variables in the latter sections. In the second part (Section 3 – Design Evaluation), we demonstrate how material properties are evaluated from design variables using the proposed model. This demonstration is then followed by Section 4 – Design Synthesis, in which Gaussian Process metamodeling is used to create a surrogate response surface between the latent variables and the objective property/performance, and a GP-Hedge Bayesian optimization is applied to optimize the microstructure to achieve the target material property. After that, we elaborate two additional features of the proposed methodology -- scalability which provides flexibility in taking arbitrary sized input/output, and transferability which makes it possible to utilize the trained weights to build a more accurate structure-property predictive model (Section 5). Last but not the least, we draw conclusions and discuss potential directions to further extend this proposed methodology.
\section{Microstructural Design Representation using Deep Adversarial Learning}
In the proposed methodology, the deep adversarial learning approach, specifically Generative Adversarial Network (GAN), is first used to identify a set of latent variables as microstructure design variables based on microstructure images collected for the same material system. In this section, the fundamentals of GAN are first introduced. It is then followed by a presentation of the proposed network architecture and designated loss function. Finally we specify some training details of the proposed deep adversarial learning model.
\subsection{Fundamentals of Generative Adversarial Network (GAN)}
Generative Adversarial Network is a type of deep generative neural network first proposed by Goodfellow et al. \cite{goodfellow2014generative, makhzani2015adversarial}. Originated from game theory, the training process of GAN is essentially a two-player competitive game. Specifically, GAN trains a generator network $G(\textbf{z};\mathbf{\theta}^{(G)})$ that produces samples $x_{G}$ from latent variables $z$ to approximate real samples $x_{data}$, and a discriminator network $D(x)$ that distinguishes the generated samples from the real samples. This competitive game would eventually lead to a Nash Equilibrium \cite{osborne1994course} between the generator $G$ and the discriminator $D$. A more vivid analogy of GAN is given by Goodfellow et al. \cite{goodfellow2014generative}: in this adversary scenario, the generator can be thought of a group of counterfeiters who tries to produce fake currency, while the discriminator is analogous to a team of police, trying to detect the counterfeit currency from the real money. Competitions in this adversary game would keep pushing both sides to the equilibrium in which the counterfeits are indistinguishable. When the generator is capable of producing realistic samples at the equilibrium, the latent variables $z$ would be naturally taken as the ``code'' of the data. In the context of proposed generative microstructural design framework, the ``code'' will serve as the design variables to create new microstructure designs.\\
\begin{figure}[!ht]
\centering
\includegraphics[width=8.5cm, height=8cm]{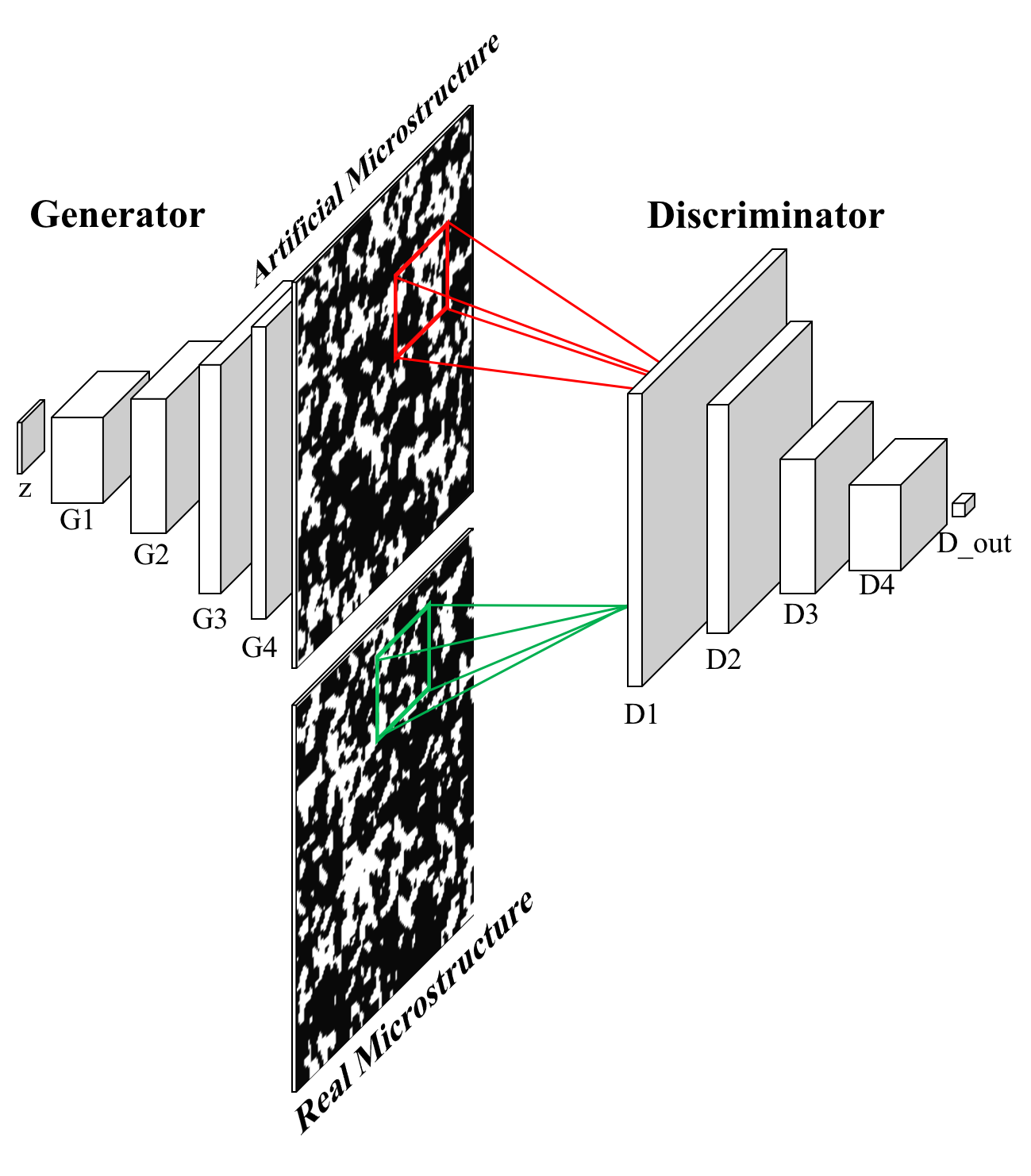}
\caption{The architecture of the proposed generative adversarial network.}
\label{figure_GAN}
\end{figure}
\indent An illustration of GAN is shown in Figure~\ref{figure_GAN}. The latent variable space is denoted as $Z$ while the microstructure data space is represented by $X$. On the left hand side, to learn the generator distribution $p_g$ that approximates the data distribution, a prior distribution of the latent variables is defined by $Z\sim p_{\textbf{z}}(\textbf{z})$. $\textbf{z}$ is then propagated through a deep neural network to create a differentiable mapping $G(\textbf{z};\mathbf{\theta}^{(G)})$ from the latent variable space $Z$ to microstructure data space $X$. On the right hand side, we also define a discriminator network that takes $\textbf{x}$, either generated or real microstructures, and produces a scalar label that indicates if $\textbf{x}$ is from real data. In other words, we train discriminator $(D)$ to maximize the probability of assigning the correct label to both real (label=1) and generated samples (label=0), while we train generator $(G)$ to maximize the number of occurrences that the labels are incorrectly assigned by $D$. Essentially, $D$ and $G$ plays a two-player minimax game, which can be expressed as the following equation:
\begin{equation}
\begin{multlined}
\min_{G} \max_{D} V(D,G) =
\E_{X\sim{p_{data}(x)}}[\log{D(\textbf{x})}] \\+ \E_{\textbf{z}\sim{p_z(z)}}[\log(1-D(G(\textbf{z})))]
\end{multlined}
\label{minmax_GAN}
\end{equation}
\subsection{Network Architecture }
In this work, the architecture of deep convolutional generative adversarial network in  \cite{radford2015unsupervised} is adopted except that we use convolutional layers to replace the fully-connected layers in both generator and discriminator for the sake of scalability (this will be introduced in section 5). The generator and the discriminator have the same number of layers, and the number of the convolutional filters are aligned symmetrically in the generator and the discriminator. In the generator, the last de-convolutional layer is associated with a $\tanh$ activation function to produce images with bounded pixel values, while the other de-convolutional layers are attached with batch normalization operations \cite{ioffe2015batch} and Rectified Linear Unit (ReLU) activations \cite{nair2010rectified}. In the discriminator, the last convolutional layer has a sigmoid activation function appended to produce probabilities between 0 and 1, while the other convolutional layers are all associated with batch normalization operations \cite{ioffe2015batch} and leaky Rectified Linear Unit (Leak ReLU) activations \cite{maas2013rectifier}. \\
\indent Figure~\ref{figure_GAN} is a simple demonstration of the proposed architecture with 5 layers in both generator and discriminator. It should be noted that, arbitrary number of layers could be applied in the proposed architecture, as long as the symmetry is kept. 
\subsection{Loss Function}
While the optimality of GAN model is Nash equilibrium theoretically, in practice, the global optimality or sufficiently good local optimality is not guaranteed \cite{salimans2016improved}. A common example of failure is the model collapse, in which the generator converges to a state that consistently produces identical samples. Therefore, in order to produce morphologically and statistically equivalent microstructures from the generator, we carefully design the loss function which can be generalized to different applications (Section 2.3) and training parameters (Section 2.4). Specifically, the total loss consists of three major components: 1) {\em adversarial loss (aka. GAN loss)} that combinatorially evaluates the performance of generator and discriminator, 2) {\em Style transfer loss} that imposes morphological constraints to the generated micorstructures, and 3) {\em Model collapse loss} that prevents the training from collapsing. \\
\indent \textbf{GAN Adversarial Loss}: The GAN Adversarial Loss is essentially the optimization objective in the vanilla version of GAN (Eq.~\ref{minmax_GAN}), expressed as 
\begin{equation}
\begin{multlined}
L_{GAN} =
\E_{X\sim{p_{data}(x)}}[\log{D(\textbf{x})}] \\+ \E_{\textbf{z}\sim{p_z(z)}}[\log(1-D(G(\textbf{z})))]
\end{multlined}
\label{GAN_loss}
\end{equation}
Note again that, in the min-max training $\min_{G}\max_{D}L_{GAN}$ essentially wants the generator $G$ to minimize this loss and let $D$ maximizes it. In practice, we follow \cite{salimans2016improved} to alter the loss of $\min{(\log{(1-D)})}$ to $\max{(\log{D})}$ when optimizing $G$.

\textbf{Style Transfer Loss:} This loss essentially imposes morphology constraints to the generated samples. The style transfer loss, namely Gram-matrix loss, is originated from a work by Gaty et al. \cite{gatys2015texture} for the purpose of texture synthesis. In the field of material science, Cang et al. \cite{cang2018arxiv} included the style transfer loss into the total loss function as a penalty term when training a Variational Auto-Encoder network \cite{makhzani2015adversarial}. In our early work, Li et al. \cite{li2018transfer} takes the style transfer loss as an optimization objective and uses its gradients with respect to each entry in the microstructure image to reconstruct statistically equivalent microstructures. They also discover an interesting intrinsic relationship between the layers included in the calculation of style transfer loss and the reconstructed microstructure: higher level convolutional layers could be dropped to reduce the computational cost while preserving the reconstruction accuracy. Recognizing this intrinsic relationship, in this work, we only retain the first four lowest convolutional layers in the VGG-16 model \cite{simonyan2014very} and compute their Gram-matrix as the style representations. The style transfer loss \cite{gatys2015texture} can be expressed as 
\begin{equation}
L_{style} = \sum_l\sum_{i,j}\frac{1}{4N_l^2M_l^2}(G_{ij}^{l}-A_{ij}^l)^2 \indent (l=1,2,3,4)
\label{ST_loss}
\end{equation}
which measures the distance between style representations of generated images and real images. In Eqn.~\ref{ST_loss}, $N_l$ and $M_l$ are number of feature maps and size of each feature map (i.e. $height \times width$) of the $l^{th}$ convolutional layer. $G^l$ and $A^l$ are the Gram-matrix of generated images and real images, respectively. The formula of Gram-matrix is 
\begin{equation}
G_{ij}^l = \sum_{k}F_{ik}^lF_{jk}^l
\end{equation}
which calculates the inner product between the $i^{th}$ and $j^{th}$ vectorized feature maps of the $l^{th}$ convolutional layer. 

\textbf{Model Collapse Loss:} Model collapse is a common problem of training a GAN model where the generated samples are clustered in only one or few modes of $p_{data}(x)$. Thus, model collapse loss \cite{zhao2016energy}
\begin{equation}
L_{collapse} = \frac{1}{n(n-1)}\sum_{i}\sum_{j\neq{i}}(\frac{S_i^{T}S_j}{||S_i|| ||S_j||})^2
\label{collapse_loss}
\end{equation}
is introduced to prevent the training from getting into collapse mode. In this equation, $n$ denotes the number of samples in a batch and $S$ represents a batch of sample representations from outputs of the first four convolutional layers of VGG-16 model \cite{simonyan2014very}. In other words, $S$ is the concatenated vectorized feature maps of the first four convolutional layers of VGG-16 model \cite{simonyan2014very}.

\begin{figure}[!ht]
\centering
\includegraphics[width=8.5cm, height=8cm]{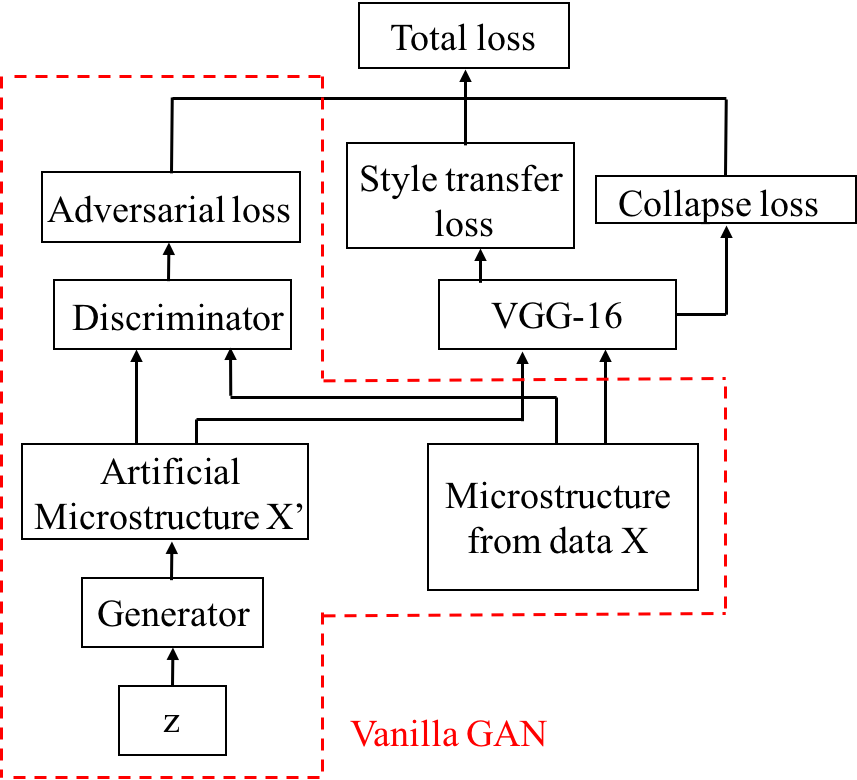}
\caption{The composition of loss function and information flow in the proposed architecture.}
\label{figure_GANflow}
\end{figure}

\textbf{The total loss:} The  total loss is a weighted combination of the three aforementioned losses.
\begin{equation}
\begin{multlined}
L(G, D) = L_{GAN} + \alpha{L_{style}} + \beta{L_{collapse}}
\end{multlined}
\label{total_loss}
\end{equation}
$\alpha$ and $\beta$ are the moderating weights that prevent the style transfer loss and model collapse loss from diminishing to zero or overwhelming the GAN adversarial loss. The composition of loss functions and the information flow in the proposed neural network architecture is depicted in Figure~\ref{figure_GANflow}.
\subsection{Numerical Validation of Latent Variables}
We apply the proposed deep adversarial learning approach to determine the latent variables for a dataset of material microstructures.\\

\begin{figure}[!ht]
\centering
\includegraphics[width = 4.5in, keepaspectratio=true]{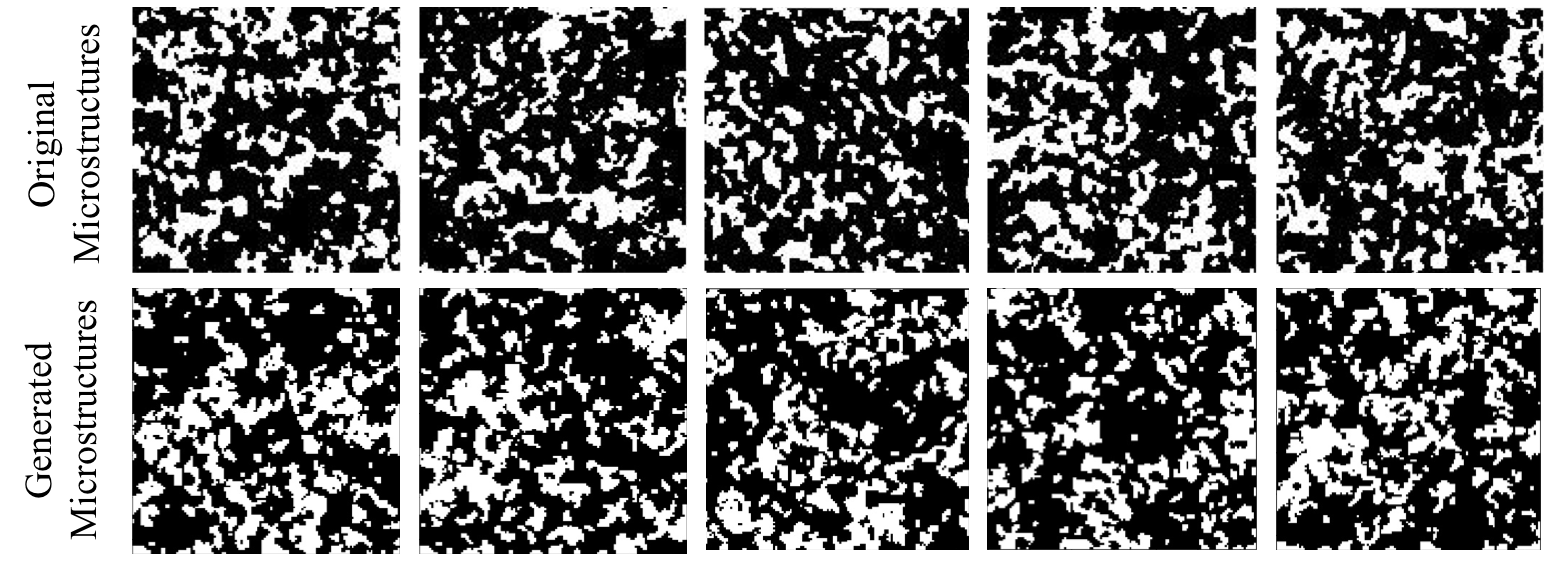}
\caption{Examples of original (training) microstructures and microstructures produced by the generator.}
\label{figure_ORGRECON}
\end{figure}

\textbf{Training data}
\indent To train the proposed GAN model, a dataset of material microstructure images that covers a variety of microstrucural dispersions are required. In addition, it is also required that all the training microstructure images share the same size. In this work, to validate the proposed approach, 5,000 synthetic microstructure images of size $128\times128$ are created using Gaussian Random Field (GRF) method \cite{jiang2013efficient}. In order to reasonably cover the vast space of compositional and dispersive patterns that correspond to different processing conditions for the same material system, three parameters (mean, standard deviation and volume fraction) are carefully controlled in the GRF model to produce microstructures with different dispersive status but sharing similar underlying characteristics of morphology. Figure~\ref{figure_ORGRECON} row 1 demonstrates some examples of the training microstructures. 5,000 of these samples are used for training the GAN model. While 5,000 seems to be an unrealistic number in material data gathering, we note that multiple images can be cropped from one microstructure image in practice. For example, for $1,000\times1,000$ sized microstructure imagess, thousands of $128\times128$ samples can be cropped with partial overlapping of the samples.

\textbf{Network architecture specifications and training parameters}
\indent One advantage of the proposed methodology is the flexibility in assigning the dimension of latent variables. The generator network is essentially a mapping between latent variables and microstructure images, so the neural network architecture depends on both the dimensionality of latent variables pre-specified and the size of microstructure images. Typically, lower dimensionality is desired for latent variables from the microstructural design perspective, because smaller number of design variables helps to reduce the computational cost in microstructure optimization. However, smaller dimensionality of latent variables will increase the depth of neural network or increase the stride parameter in the convolutional layers, which makes the training of GAN more difficult. Therefore, a trade-off between the latent variables’ dimensionality and the training difficulty needs to be considered. After several experiments, it is discovered that the 5-layer architecture with stride $2\times2$ as illustrated in Figure~\ref{figure_GAN} is practically easy to stabilize and converge while providing sufficiently low dimensionality for the latent variables. The $2\times2$ stride configuration essentially results in a scaling factor of 2 on each dimension in each layer, thus 5 stacked layers would scaling down the microstructures by a factor of $32$ (i.e. $2^5$) on each dimension. For the aforementioned dataset, the $128\times128$ images are converted to a $4\times4$ latent variable tensor, which is flattened to a 16-dimensional latent variable vector $z$.\\
\indent   In addition to the dimensionality of $z$, a bounded latent variable space is defined by setting each entry of $z$ to be independent and uniformly distributed between -1 and 1. For generator network, four (de-convolutional)-batch normalization-ReLU layers are appended to $z$ sequentially, which is then followed by a (de-convolutional)-tanh layer to produce $128\times128\times1$ sized microstructure images.   In contrast, the discriminator network is composed by four sequentially connected convolution-batch normalization-leaky ReLU layers. A convolutional-sigmoid layer is appended to the end of the discriminator network to produce a scalar valued between 0 and 1 to represent the probability of classifying if the image given to the discriminator is from real microstructure dataset (instead of artificially generated ones). A detailed specification of the dimensionality in each layer is illustrated in Table ~\ref{table_netarch}. Note that to achieve the specified dimensionality, in both de-convolutional and convolutional layers, the filter size is set as $4\times4$ and strides are all $2\times2$ (The only exception is that we use $8\times8$ filter with stride $1\times1$ between discriminator layer 4 and 5).\\
\begin{table}[h!]
\centering
\caption{The dimensionality of each layer in the proposed network architecture. (bs. is the abbreviation of batch size)}
\begin{center}
\label{table_netarch}
\begin{tabular}{c l l}
\hline
Layer &Dimension \\
\hline
Random Tensor z & $bs.\times4\times4\times1$ \\
Generator Layer 1 & $bs.\times8\times8\times128$ \\
Generator Layer 2 & $bs.\times16\times16\times64$ \\
Generator Layer 3 & $bs.\times32\times32\times32$ \\
Generator Layer 4 & $bs.\times64\times64\times16$ \\
Image X & $bs.\times128\times128\times1$\\
Discriminator Layer 1 & $bs.\times64\times64\times16$\\
Discriminator Layer 2 & $bs.\times32\times32\times32$\\
Discriminator Layer 3 & $bs.\times16\times16\times64$\\
Discriminator Layer 4 & $bs.\times8\times8\times128$\\
Discriminator Layer 5 & $bs.\times1\times1\times1$\\
\hline
\end{tabular}
\end{center}
\end{table}
\indent The $\alpha$ and $\beta$ parameters discussed in Section 2.3 are set as 0.03 and 0.03 for optimal balance between the three components of losses, respectively. Adam optimizer \cite{kingma2014adam} is applied in training by setting the learning rate as 0.0005, $\beta_1$ value as 0.5 and $\beta_2$ value as 0.99. In the alternating training of the generator $G$ and the discriminator $D$, it is found that it is optimal to set the ratio of network optimization for discriminator and generator to 3:1 (i.e. update discriminator three times and then update generator once) to achieve stability and convergence.\\
\indent Some other significant training parameters include: number of epochs -- 15,000; batch size -- 30 and the $\alpha$ parameter in leaky ReLU -- 0.2.\\

\begin{figure}[!ht]
\centering
\includegraphics[width = 4.5in, keepaspectratio=true]{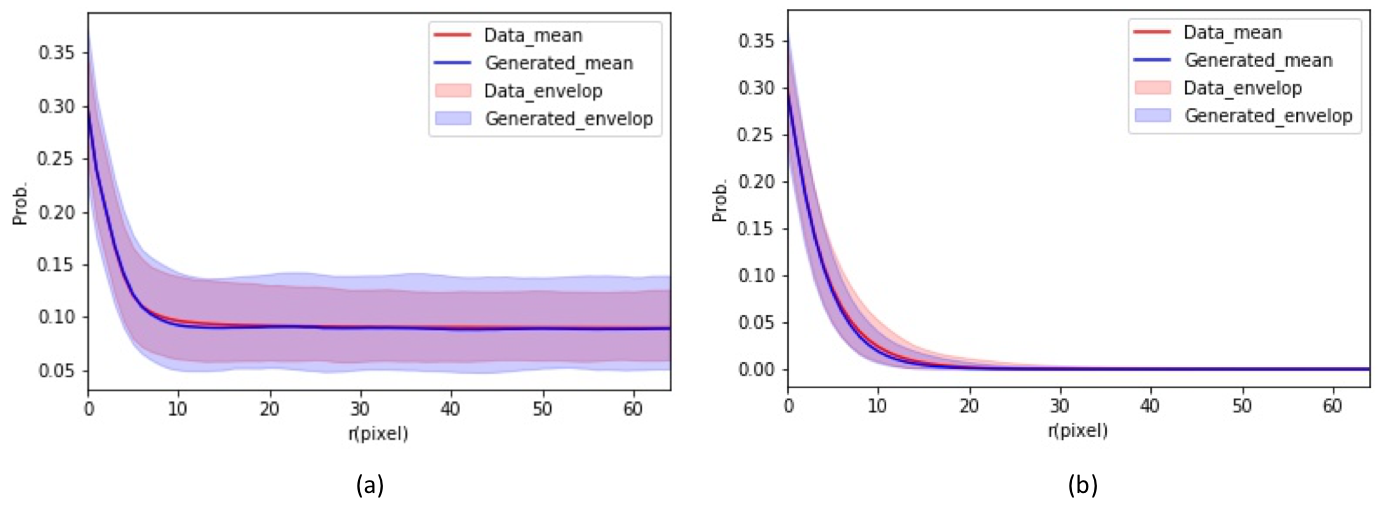}
\caption{Comparison of correlation functions of original microstructures and microstructures generated by the proposed generator.(a) Two-point correlation function. (b) Lineal-path correlation function.}
\label{figure_twop}
\end{figure}

\textbf{Validation of the latent variables}
\indent The validity of the latent variables and the amount of information loss are evaluated by comparing the original microstructure set and a set of microstructures produced by randomly sampling latent variables $z$ and propagating through the generator network. Specifically, we compare the two-point correlation functions \cite{jiao2007modeling,ramin2018review} and lineal-path correlation functions \cite{lu1992lineal} of the 5,000 original microstructures and 5,000 generated ones produced by the generative model trained in GAN. Figure~\ref{figure_ORGRECON} shows that the generator in GAN is capable of producing visually similar microstructures as the original image data used for training.  Figure \ref{figure_twop} shows the two-point and lineal-path correlation functions of original microstructures and microstructures generated by the proposed generator. Figure~\ref{figure_twop} (a) and (b) show that the mean correlation functions of the 5,000 training samples matches those of the 5,000 generated ones. In addition, the two-point correlation functions’ envelop of the generated samples overlaps with all possible regions that the original data covers and its slightly broadened envelop suggests that the proposed model might be capable of extrapolating the range of microstructures (by exploring more possibilities of the microstructures) while retaining the morphological characteristics of the collected samples.
\section{Microstructure Design Evaluation}
In the context of microstructural materials design, design evaluation is the process of evaluating the material properties of interest for a generated microstructure controlled by the design variables. In the proposed methodology, it includes two steps: 1) Latent variables (design variables) to microstructures: the GAN generator learned in the deep adversarial learning is used to propagate the values of latent variables to obtain microstructure images. 2) Microstructure to material property: For a generated microstructure, physics-based simulation is used to obtain the corresponding material property or structure performance. For the case study in this work, the Rigorous Coupled Wave Analysis (RCWA)~\cite{yu2017characterization} is used to simulate the optical absorption performance of the given microstructure.
\section{Microstructure Design Synthesis}
Each entry of the latent variables vector $z$ identified by GAN is independent and bounded in [-1 1]. They serve as the microstructure design variables in design synthesis which is accomplished through simulation-based optimization. Since the structure-property or structure-performance evaluation is often computationally expensive, a Bayesian optimization approach is applied to search for the optimal microstructure with desired material behavior through sequential adaptive sampling. The design optimization problem is formulated as

\begin{equation*}
\begin{aligned}
& z=\operatorname*{argmin}_z -f(G(z))\\
& \text{s.t.\indent}  z_i \in [-1,1].
\end{aligned}
\label{designformulation}
\end{equation*}
where $G(\cdot)$ is the generator mapping in GAN, and $f(\cdot)$ is the physical simulation. After obtaining the optimal value of $z$, the optimal microstructure can be generated rapidly by generator $G(z)$.\\
\indent In the remaining part of this section, we illustrate the use of response surface-based Bayesian optimization through a microstructural materials design case study. The 2D metamaterial structures being explored have similar morphological characteristics as the ones used in Section 2 (Figure~\ref{figure_ORGRECON} row 1), but a smaller size of $96\times96$ pixels. The design objective is to obtain the microstructure that maximizes the optical absorption simulated by RCWA, a desirable performance in applications such as solar cell design.  The learned model in Section 2 is applied in this case study, and the dimensionality scaling factor is still $\times32$ in each dimension. In other words, the $96\times96$ microstructure images would be represented by $3\times3$ dimensional tensor (i.e. 9-dimensional vector).

\subsection{Exploration of Design Variable Space using Design of Experiments (DoE)}
To create the response surface model between the design variables and the objective material property, a set of design of experiments (DOE) are sampled. In this work, Latin Hyper-cube Sampling (LHS) \cite{mckay2000comparison} is applied to sample 250 points in the 9-dimensional space. Then the material optical performance for these designs, denoted as $y$, is obtained by following the design evaluation process described in Section 3. The dataset of 250 samples $(\mathbf{z}, y)$ are used to create the initial response surface model for Bayesian optimization.
\subsection{Gaussian Process Metamodeling and GP-Hedge Bayesian Optimization}
\begin{figure}[!ht]
\centering
\includegraphics[width = 4.5in, keepaspectratio=true]{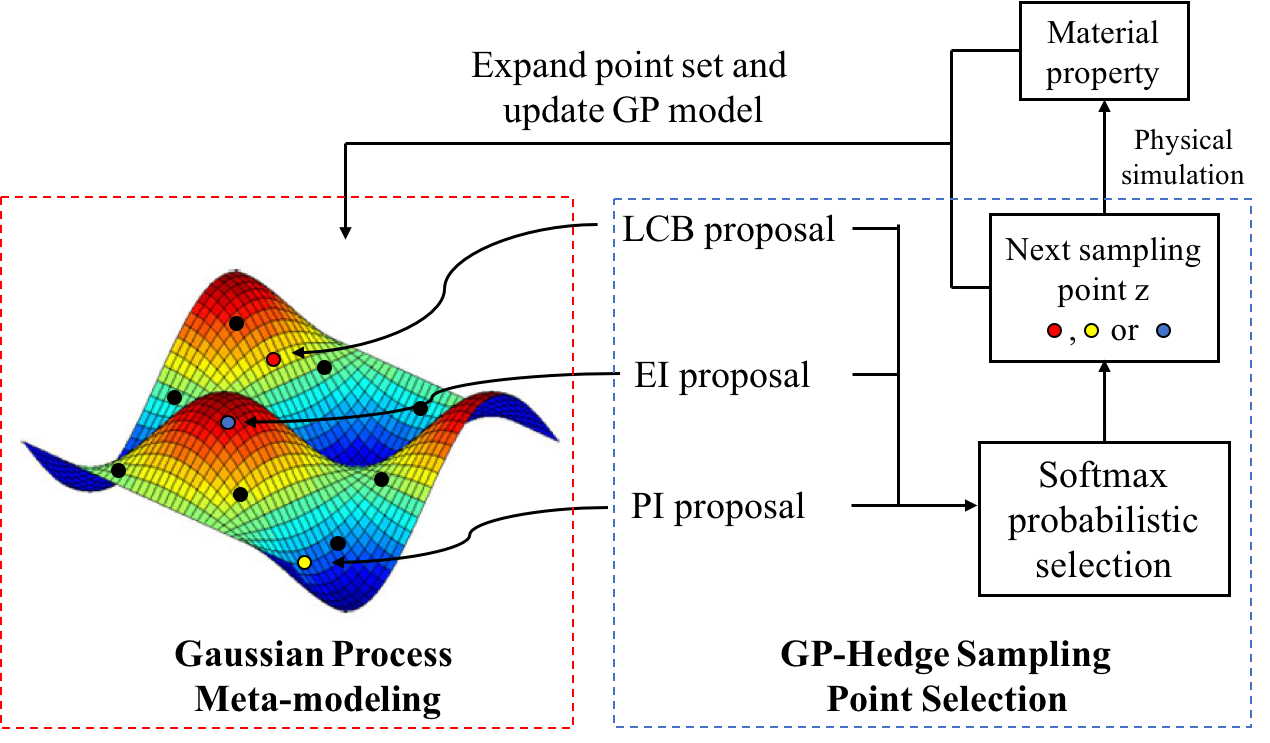}
\caption{The integration of Gaussian process metamodeling and GP-Hedge Bayesian optimization.}
\label{figure_gphedge}
\end{figure}

After the initial sampling using LHS, metamodel-based Bayesian optimization is conducted to iteratively explore the potentially optimal design point. Compared to stochastic optimization approaches such as Genetic Algorithm (GA) and Simulated Annealing (SA), Bayesian optimization is a much more efficient global optimization technique as it encourages both exploration and exploitation in the optimization search process. In each optimization iteration, we fit a metamodel (aka. surrogate model or response surface model) using \textit{Gaussian Process metamodeling} \cite{rasmussen2004gaussian} to statistically approximate the relationship between design variables and the design performance. The dataset $(\mathbf{z}, y)$ is expanded by one more sampling point in each iteration using the \textit{GP-Hedge} criteria \cite{hoffman2011portfolio}. Figure~\ref{figure_gphedge} illustrates how Gaussian Process metamodeling and the GP-Hedge optimization strategy are integrated in this work.\\ 

\begin{figure}[!ht]
\centering
\includegraphics[width = 4.5in, keepaspectratio=true]{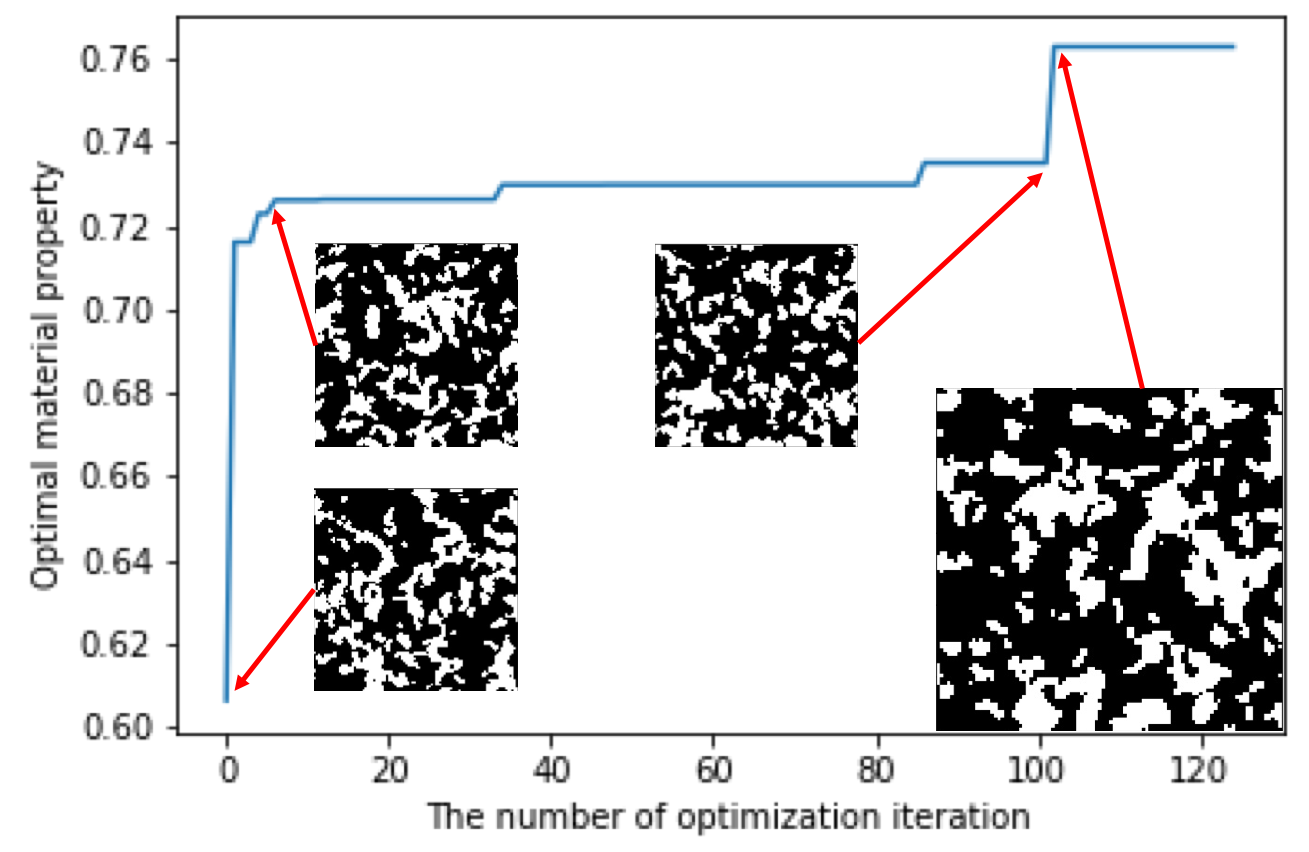}
\caption{The microstructure optimization history and microstructure designs indicated at selected iterations.}
\label{figure_designEvol}
\end{figure}

\begin{figure}[!ht]
\centering
\includegraphics[width = 4.5in, keepaspectratio=true]{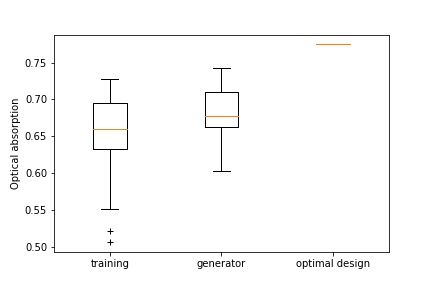}
\caption{The comparison of the optical absorption property between 1) 30 randomly generated microstructures, 2) 30 microstructures generated by the trained generator, and 3) optimal design.}
\label{figure_design}
\end{figure}

\indent Gaussian Process model \cite{rasmussen2004gaussian}, also known as Kriging model, is a statistical model that interpolates the observations and supplies uncertainty for the metamodel prediction at each estimation point. In essence, Gaussian Process models the data points $\{\mathbf{X}, \mathbf{y}\}$ and the estimations $\{\mathbf{X'}, \mathbf{y'}\}$ using
\begin{equation}
\begin{multlined}
\newcommand{\Cov}{\operatorname{Cov}}
\begin{bmatrix}
\mathbf{y} \\ \mathbf{y'}
\end{bmatrix}\sim \mathcal{N} \bigg( \mathbf{0}, 
\begin{bmatrix}
\Cov(\mathbf{X}, \mathbf{X}) & \Cov(\mathbf{X}, \mathbf{X'}) \\
\Cov(\mathbf{X'}, \mathbf{X}) & \Cov(\mathbf{X'}, \mathbf{X'})
\end{bmatrix}\bigg)
\end{multlined}
\label{GPdistribution}
\end{equation}
where Cov$(\mathbf{A},\mathbf{B})$ represents the covariance matrix between $\mathbf{A}$ and $\mathbf{B}$, defined by Cov$(\mathbf{A}, \mathbf{B})=\E(\mathbf{AB}^{T})-\E(\mathbf{A})\E(\mathbf{B})^{T}$.
Conditioning on the data $D=\{\mathbf{X}, \mathbf{y}\}$, the posterior $P(\mathbf{y'}|\mathbf{X},\mathbf{X'},\mathbf{y})$ yields a Gaussian distribution in which,
\begin{equation}
\begin{split}
\label{GPmeanvar}
&\newcommand{\Cov}{\operatorname{Cov}}
\mathbf{\mu}=\Cov(\mathbf{X}, \mathbf{X'})\Cov(\mathbf{X}, \mathbf{X'})^{-1}\mathbf{y}\\
&\newcommand{\Cov}{\operatorname{Cov}}
\Sigma=\Cov(\mathbf{X'}, \mathbf{X'})-\Cov(\mathbf{X}, \mathbf{X'})\Cov(\mathbf{X}, \mathbf{X})^{-1}\Cov(\mathbf{X'}, \mathbf{X})
\end{split}
\end{equation}
Gaussian Process metamodeling essentially gives a surrogate model that quantifies the statistical mean estimations and uncertainties at the unexplored design points. By using the mean estimations and the uncertainties, a smaller set of design points that could potentially improve the performance can be identified. In this case, expensive design evaluations only need to be conducted on these candidate design points, thereby eliminating redundant design evaluations. As a consequence, the overall computational cost of the design process is reduced tremendously.\\
\indent In each iteration of the Bayesian optimization, the Gaussian Process metamodel is applied to determine the next sampling point. Typical criterion (aka. acquisition functions) to locate the next sampling point include expected improvement (EI) \cite{jones1998efficient}, probability of improvement (PI) \cite{kushner1964new} and lower confidence bound (LCB) \cite{auer2002using}. These criterion are different in how the trade-off is made between exploration (picking samples at locations with large uncertainty) and exploitation (choosing samples at locations close to the optimum based on the mean prediction). In this work, we apply the GP-Hedge mechanism to probabilistically choose one of the above three acquisition functions at every optimization iteration. The general procedure of GP-Hedge Bayesian optimization is illustrated in Algorithm ~\ref{algo1}.  
\begin{algorithm}
\caption{GP-Hedge Bayesian Optimization}
\begin{algorithmic}[1]
\STATE Select parameter $\eta \in \mathbb{R}^+$ 
\STATE Set the gains for acquisition function $i$, $g_{0}^{i}=0$ for $i=1,2,...,N$
\STATE $t=0$
\WHILE{stopping criteria is not met}
\STATE $t=t+1$
\STATE Each acquisition function propose a point $\mathbf{x}_{t}^{i}$
\STATE Set $\mathbf{x}_t=\mathbf{x}_t^{i}$ with softmax probability $p_t^{i}=\exp{(\eta g_{t-1}^{i})}/\sum_{l=1}^{K}\exp{(\eta g_{t-1}^{l})}$
\STATE Obtain the objective function value $y_t=f(\mathbf{x}_t)$
\STATE Augment data $D_{1:t} = \left\{{D_{1:t-1}, (\mathbf{x}_t,y_t)}\right\}$
\STATE Receive rewards $r_t^{i}=\mu(\mathbf{x_t^{i}})$ from the updated GP
\STATE Update gains $g_t^{i} = g_{t-1}^{i}+r_{t}^{i}$
\ENDWHILE
\end{algorithmic}
\label{algo1}
\end{algorithm}
This GP-Hedge Bayesian optimization process is applied to our design case study beginning with the metamodel created using the 250 initial LHS samples, followed by 120 iterations of optimization. Throughout the optimization process, the values of the latent (design) variables are constrained between [-1, 1] to retain the morphological characteristics learned from the sample images.\\
\indent Figure~\ref{figure_designEvol} shows the optimization history with microstructure design solution indicated at a few iterations.  A few observations can be made:  1) Design performance is improved significantly at the very beginning of the Bayesian optimization, while the improvement becomes less as the number of iterations increases. 2) Design performance is not necessarily improved in the new iteration. This is reasonable because the new sampling point is chosen for both exploration and exploitation using the criterion that combines both the mean estimation and the uncertainty in the metamodel.\\
\indent Figure~\ref{figure_design} illustrates the comparison between the optical performance of three datasets: a) 30 randomly sampled microstructures from training set, b) 30 microstructures generated by randomly sampling latent variables $z$ and propagating through the trained generator, and c) the optimized microstructure. It should be noted that in order to make a fair comparison, we randomly sampled 30 microstructures from training set in each trial for dataset a), and repeated this trial 10 times. It is observed that the results of randomly sampled microstructures have the lowest optical performance and the largest variance. It is found that the mean optical performance of the microstructures produced by the GAN generator (0.6827) is 4.8\% ($0.6827/0.6509-1$) greater than that of the randomly sampled microstructures (0.6509), while the optimized microstructure's performance (0.7630) exceeds the mean performance of randomly sampled microstructures by 17.2\% ($0.7630/0.6509-1$). It should be noted that the theoretical upper bound of the evaluated optical absorption property is $1.0$, so the design solution provided by the proposed approach is reasonably good. These results verify the effectiveness of the proposed design optimization framework.
\section{Scalability and Transferability}
In the previous sections, we have discussed the process of applying the proposed deep adversarial learning model for identifying latent variables of microstructures and conducting microstructural materials design. With the proposed methodology, the dimensionality of latent (design) variables can be prescribed and the information loss is negligible even for complex microstructural geometries. In addition to these advantages, in this section a few additional useful features of the proposed deep adversarial learning model are elaborated.
\subsection{Scalability of the generator}
\begin{figure}[!ht]
\centering
\includegraphics[width = 4.5in, keepaspectratio=true]{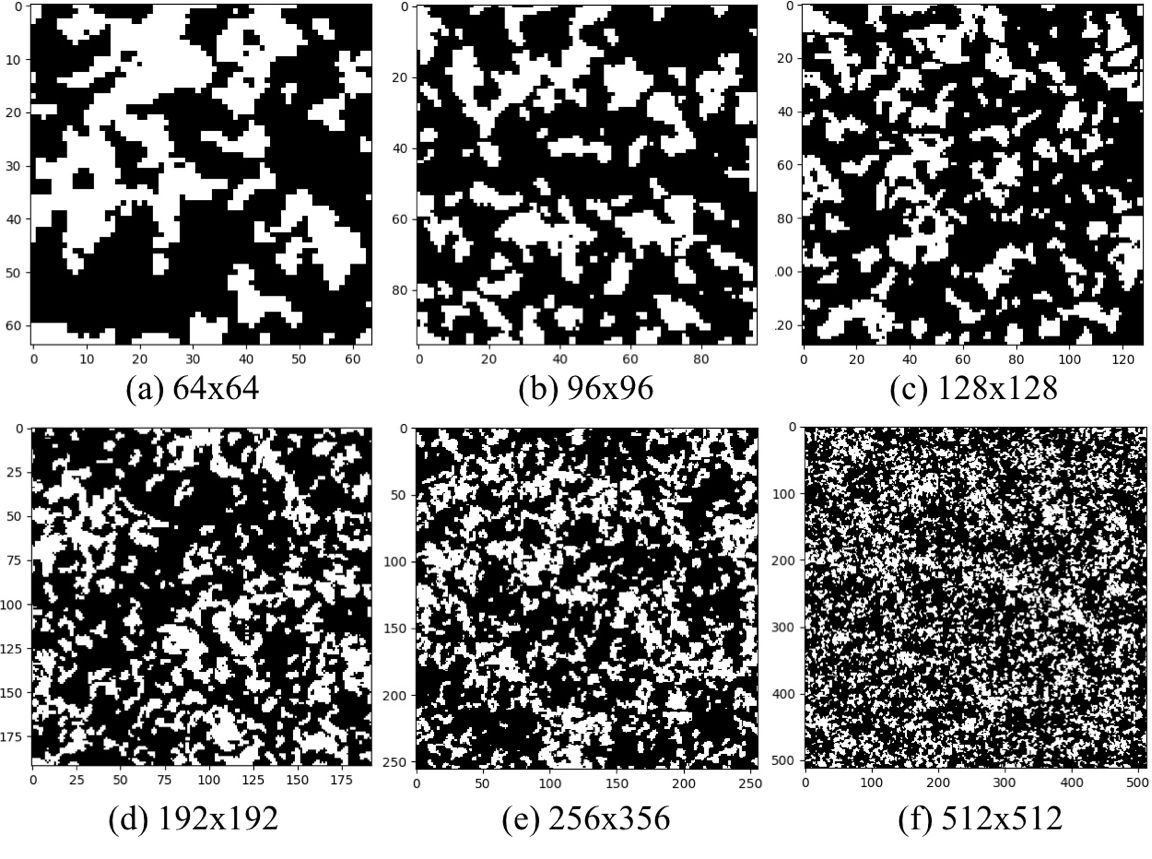}
\caption{An illustration of microstructures of different sizes generated by the scalable generator.}
\label{figure_scalableG}
\end{figure}

\indent Benefited from the exclusion of fully-connected layers in the network architecture, the scalability of the generator provides the proposed GAN model the flexibility of taking arbitrary sized inputs (latent variables) and outputs (microstructures). This is a signature of the proposed model because confining the input dimensionality could lead to a low dimensional microstructural design space, and varying the output size can consequentially produce different sized microstructures to serve different analytical purposes (e.g. analysis in Statistical Volume Elements (SVEs) vs. Representative Volume Elements (RVEs)). 
\\
\indent Specifically, the scalability is useful in two ways:  \textbf{a) Flexibility in setting the dimensionality of latent variables.} In the proposed network architecture, adding each additional convolutional layer increases the scaling factor between the generated image and the latent variables by a factor of 4 ($\times2$ on each dimension). Therefore, in the aforementioned design case in Section 4, when the 9216-dimensional ($96\times96$) microstructure is to be converted into 9-dimensional ($3\times3$) latent variables, five network layers are stacked (i.e. $96/3=32=2^5$). In theory, stacking more neural network layers in the proposed model can enlarge the scaling factor, and the accuracy would be retained as long as the training is well handled. However, adding more layers inevitably increase the difficulty of training the GAN. In other words, while low dimensionality of the latent variables often leads to less microstructure design optimization cost because of less design variables, it increases the GAN training cost because of higher model complexity. Hence a key consideration in choosing the number of latent variables is the trade-off between the optimization cost and the GAN training cost. When the computational resource for design optimization is limited (e.g. physics-based simulations are extremely expensive), it would be better to keep a lower dimensionality of the latent variables though more training time for GANs is needed. In contrast, if the design optimization is not limited by the computational resource, a reasonably higher dimensionality of the latent variables is acceptable so that the burden on training GANs can be reduced. \textbf{b) Generating arbitrary sized microstructures.} While the deep learning network is trained by setting the dimension of $z$ as $bs.\times4\times4\times1$, one may modify the dimensionality of latent variables $z$ to control the size of the generated images without retraining the model. Figure~\ref{figure_scalableG} illustrates the generated images with different sizes using different dimensional settings of $z$. It demonstrates that the proposed generator is capable of generating arbitrary sized microstructures for the material system of interest. An alternative way of controlling the size of microstructure is to include/remove convolutional layers. For instance, $256\times256$ images could be generated by adding one more layer in both generator and discriminator before training and keep the size of $z$ as $bs.\times4\times4\times1$ However, the deeper the neural network is, the harder the training process would be. Moreover, retraining is required if model's architecture is changed. Hence changing the dimensionality of $z$ is often preferred for this reason.

\subsection{Transferability of the discriminator}
\begin{figure}[!ht]
\centering
\includegraphics[width = 4.5in, keepaspectratio=true]{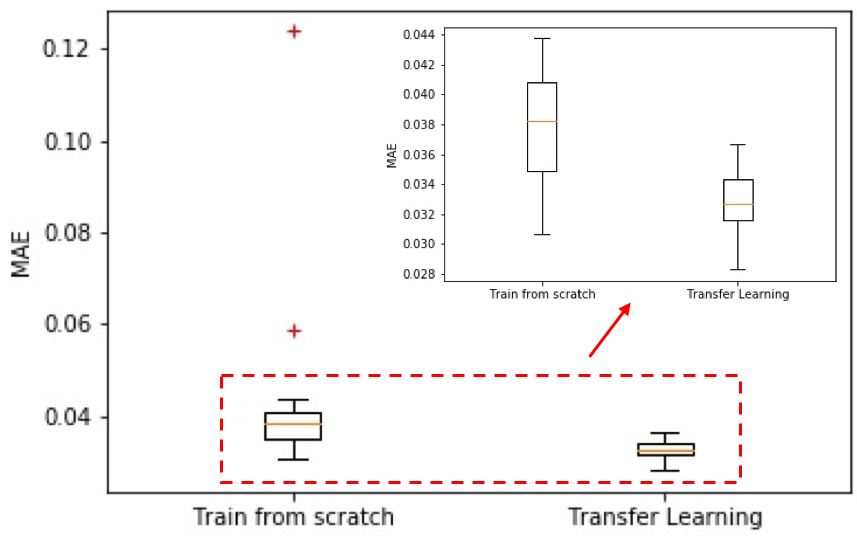}
\caption{The comparison of the mean-absolute-error (MAE) for training from scratch and transfer learning. Outliers, higher error and variance are observed from the results of ``training from scratch'' control group.}
\label{figure_mae}
\end{figure}

In addition to the aforementioned microstructural materials design contributions of the proposed approach, we also discover an additional utility of the discriminator in improving structure-property predictions via transfer learning. While the generative capability is usually emphasized \cite{chen2016infogan,goodfellow2014generative}, the utilization of discriminator is more or less ignored. However, totally discarding the discriminator is wasteful as there is always significant ``knowledge'' about the data (in the context of this work, microstructures) learned by the discriminator. In this work, we propose to leverage the knowledge learned from the discriminator into the development of machine learning-based structure-property predictions via transfer learning. In training deep networks, Stochastic Gradient Descent (SGD) based algorithms are the typical choices. Since SGD converges to local minimum, its optimized value is very sensitive to the initialization of the network. With transfer learning, instead of randomly selecting a starting point for the weights of the structure-property predictive network, the weights are initialized using ones obtained in the GAN discriminator trained on the microstructure dateset in Section 2, by analogy to \cite{yosinski2014transferable}.
\\
\indent In the context of this work, the discriminator is essentially a binary classifier trained together with the generator to distinguish generated microstructure from real ones. Our objective is to utilize the trained weights in this classifier and transfer them into a structure-property regression model. It should be noted that, in training and testing this regression model, we use additional 250 samples (microstructures and their corresponding properties) exclusive from the 5,000 samples used for training the GAN, and we randomly split them into 200/50 sets for training/testing. There are three primary steps in building the regression model:
\\
\indent \textbf{1) Transferring partial architecture and weights:} We borrow the first four convolutional layers of the trained discriminator (their architecture and the corresponding weights) as the basic building blocks.\\
\indent \textbf{2) Appending full-connected layers at the end:} The output of the $4^{th}$ convolutional layer is flattened and two fully-connected layers of 2048 and 1024 neurons with ReLU activation are appended. Dropout normalization ($p=0.5$) is applied after each fully connected layer. A fully-connected layer of 1 neuron is added at the end to produce the scalar output of the regressor. The weights of all these additional layers are initialized randomly.\\
\indent \textbf{3) Fine-tuning weights using Adam:} Adam optimizer is applied to fine-tune the weights in some of the layers. As it is well recognized that the early convolutional layers (Convolutional layers 1-3 of the discriminator) usually contains general Gabor-like filters, we freeze these layers' weights from Adam optimization. The other layers are subject to the Adam optimization (learning rate=0.0005, $\beta_1$=0.5, $\beta_2$=0.99) for 4,000 epochs with batch size of 50.\\
\indent To demonstrate the advantage of applying this transfer learning strategy for building the structure-property model, we also conduct another training process with exactly the same network architecture but initializing all weights randomly (instead of using pre-trained weights) as a control group. This control group is named ``training from scratch'' in the remainder of this section. We compute the mean-squared-errors (MSE) and the mean-absolute-errors (MAE) on the 50 reserved testing data with 30 repetitive trials. Since both error metrics measure the same fundamental error phenomena, we only show the result of MAE comparison in Figure~\ref{figure_mae}. From the results, it is found that, compared to training from scratch, transfer learning strategy can facilitate the development of structure-property predictive model by improving its accuracy and stability. This finding is consistent with our intuition that prior knowledge learned by the discriminator network could help in building a more accurate predictive model. 

\section{Conclusion and Future Work}
In this work, we proposed a deep adversarial learning methodology for microstructural material design. In the proposed methodology, the dimensionality of latent variables for microstructures are prescribed first. Then a GAN consisting of a generator and a discriminator is trained on a dataset of microstructures being studied. The latent variables are then taken as design variables in a Bayesian optimization framework to obtain the microstructure with desired material property. Gaussian Process metamodeling is used at each optimization iteration to update the relationship between the design variables and the microstructure performance, and GP-Hedge criterion is used for proposing the next candidate sampling point. The proposed methodology features several contributions: First, the proposed methodology provides an end-to-end solution for microstructural materials design, which reduces information loss and preserves more microstructural characteristics. Second, this work is to extend the use of GAN to be a part of the design loop. The GP-Hedge Bayesian optimization incorporates Gaussian Process metamodeling to reduce the number of design evaluations and thus decreases the computational cost while improving the design performance. Third, a customized loss function with the proper moderating parameters is presented for generating new microstructural design with similar characteristics. Finally, the deep learning network architecture and the training parameters obtained in this work could be re-used as a starting point for other applications of deep learning in materials science (e.g. transfer learning).

\indent While this work demonstrates the benefits of the proposed methodology, a few technical details can be further examined in future work. First, this work could make a boarder impact on other material microstructures such as ones with very sharp features (e.g. pointy edges), crystalline structures or grain boundary maps, multiphase or continuous phase microstructures. Next, the processing or manufacturing constraints are not considered in the design optimization. In order to take the processing conditions as design variables, the processing-structure-property (PSP) linkage needs to be established. Similar to our earlier work \cite{hassinger2016toward, yu2017characterization, lee2017concurrent}, we will study the relationship between latent variables and such processing or manufacturing parameters, including appropriate constraints in the optimization process. Attempts would be also made to associate physical meanings to the learned latent variables so that materials scientists could explicitly control some characteristics of the optimized microstructure. In addition, the choice of dimensionality of latent variables can be guided through detailed numerical studies to better understand the impact of low dimensionality on network training. Special attention needs to be paid towards the network theory and practice for stabilizing the training process. Other potential directions for improving network modeling include but are not limited to utilizing Wasserstein GAN \cite{salimans2016improved} for solving model collapse problem, introducing ResNet structure \cite{he2016deep} for higher learning capability, or investigating visual attention mechanism \cite{mnih2014recurrent} for better interpretation of the model.

\section*{Acknowledgements}
The Rigorous Couple Wave Analysis simulation is supported by Prof. Cheng Sun's lab at Northwestern University. This work is primarily supported by the Center of Hierarchical Materials Design (NIST CHiMaD 70NANB14H012) and Predictive Science and Engineering Design Cluster (PS\&ED, Northwestern University). Partial support from NSF awards DMREF-1818574, DMREF-1729743, DIBBS-1640840, CCF-1409601; DOE awards DE-SC0007456, DE-SC0014330; AFOSR award FA9550-12-1-0458; and Northwestern Data Science Initiative is also acknowledged. 
\section*{References}


\bibliographystyle{model1-num-names}
\bibliography{reference.bib}

\end{document}